# Making Meaning with Math in Physics: A Semantic Analysis

Edward F. Redish and Ayush Gupta
Department of Physics, University of Maryland,
College Park, MD 20742-4111 USA
[redish@umd.edu; Ayush@umd.edu]

## Abstract

Physics makes powerful use of mathematics, yet the way this use is made is often poorly understood. Professionals closely integrate their mathematical symbology with physical meaning, resulting in a powerful and productive structure. But because of the way the cognitive system builds expertise through binding, experts may have difficulty in unpacking their well-established knowledge in order to understand the difficulties novice students have in learning their subject. This is particularly evident in subjects in which the students are learning to use mathematics to which they have previously been exposed in math classes in complex new ways. In this paper, we propose that some of this unpacking can be facilitated by adopting ideas and methods developed in the field of cognitive semantics, a sub-branch of linguistics devoted to understanding how meaning is associated with language.

## I. Introduction

Many sciences, particularly physics, make extensive use of mathematics. Even introductory physics texts include thousands of equations and a graduate student may find that equations cover nearly half the pages in an advanced textbook. Though the procedures and equation manipulations might appear the same, what physics does with math is dramatically different from what one learns in a math class. Physics requires us to integrate our understanding of the physical world with symbolic relations of mathematics – adding meaning and structure to both. Physics puts *meaning* to the math, adding additional levels of structure, interpretation, and even tools.

Instructors in upper division physics classes often complain that even students with satisfactory grades in their math courses "don't know enough math." Sometimes these instructors call for the students "to take more math classes," or suggest, "we should teach the math ourselves." Indeed many in the latter group feel that the math is "reasonably simple" and "could be taught in a week or two."[1] It appears that the issues of how math is used in physics is not simple and needs some careful analysis.

In order to better understand the use of math in physics so that we can teach it more effectively, we need to analyze the cognitive components of expertise and build an understanding of how it develops. Since the critical issue appears to be "making meaning" with math, we turn for help to the subject of *cognitive semantics* – a subfield of linguistics that is concerned with how ordinary language is imbued with meaning.[2] Researchers in that field are attempting to answer the question, "What do we mean by meaning?" They appear to have made considerable progress on this issue in the past twenty years and they have developed a number of concepts and a terminology that may help us deconstruct the way meaning is put to math in science.

---

[1] Personal communication over many years with many colleagues.
[2] Note that in an early era, linguists made a distinction between *semantics* – the lexical meaning of words, and *pragmatics* – the meaning of words in contexts. The linguists we follow have come to the conclusion that this is not a useful distinction and that all semantics is pragmatics.

In this paper, we argue that the barrier to effective use of mathematics in physics is not an issue of deficiency of skills or knowledge that can be solved by 'more and better instruction in math'. Using ideas from cognitive linguistics, we argue that the network of knowledge that expert physicists use to put "meaning" to mathematics is different from the ways that novices interpret mathematics in physics.

We begin by presenting some examples that illustrate how physics is added to mathematics to create new structures, ones that construct meaning beyond the mathematical. We then outline some of the most relevant basic ideas from cognitive semantics, in particular, *encyclopedic meaning* and *blending*. We then present the case of two students, both of whom are good at solving a physics problem, but who use different approaches to thinking about the math. We present a comparative analysis using the principles of cognitive linguistics to unpack the differences in the ways that they look at the same equation. Finally, we consider the implications of our analysis for research and instruction. Note that this is essentially a *theoretical* paper. Although we cite examples, they are anecdotes, not data, chosen to reify and help the reader make sense of the ideas. The value of this approach will have to be evaluated by its productivity in leading to new ideas for observations and the analysis of data.

## II. Putting Physics into the Math

### Physics models the physical world using math

When we use math in physics – or in any science – we are modeling some aspect of the physical world by specifying the way physical measurements are related and how they behave under changes of perspective (Poincare, 1905). These translations from physical-causal relations between entities in the real world to mathematical relations between symbols is what we call mathematical modeling.

In order to explicate the various components of the modeling process for the purpose of thinking about teaching mathematical physics, we use the diagram shown in *Figure 1*. (Redish 2005, 2008)

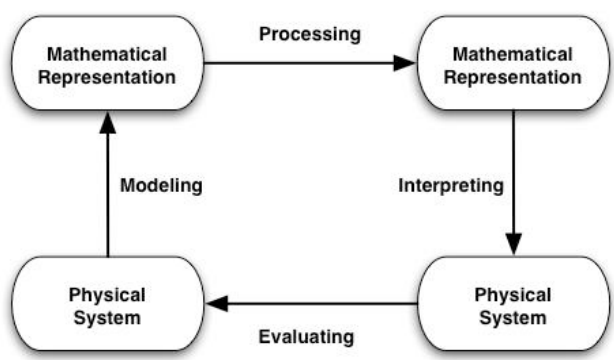

*Figure 1: A model of mathematical modeling*

We begin in the lower left with a particular physical system we want to describe. Critical here is a focus on the particular aspects of the universe we want to describe. We identify measurables –- variables and parameters we can quantify through some process of measure-ment (at least in principle, if not always in practice). We then decide what particular mathematical structures are appropriate for describing the features we are interested in modeling. We then *model* the physical system by mapping these measurements into mathematical symbols and expressing the physical-causal relations between the measured quantities in terms of mathematical operations between the symbols. From the mathematical structures we have chosen, we inherit transformational rules and methodologies for transforming relationships and solving equations. We can then *process* the math to solve problems that we were unable to see the answer to directly when we were only thinking about the physical system. This, however, leaves us with only mathematics. We then have to *interpret* what the result means back in the physical system. Finally, we have to *evaluate* whether the result supports our original choice of model when compared to observations or whether it indicates a need for a modification of our model.

All of these four steps – modeling, processing, interpreting, and evaluating – are critical skills in the toolbox of a scientist who uses math to describe the behavior of the physical world. This diagram helps a teacher focus on more than just the mathematical processing that often tends to dominate instruction in physics. But the use of math in physics is not so simple or sequential as this diagram may tend to indicate. The physics and the math get tangled.

**We re-interpret the math depending on the physics**

We not only *use* math in doing physics, we use physics in doing math. We will present three examples.

*Corinne's Shibboleth: The meaning of function attributes*

Our first example is "Corinne's Shibboleth"[1]. (Dray & Manogoue, 2002) The particular example shown in Figure 2 tends to separate physicists from mathematicians. Try it for yourself before reading the discussion that follows.

> One of your colleagues is measuring the temperature of a plate of metal placed above an outlet pipe that emits cool air. The result can be well described in Cartesian coordinates by the function
>
> $$T(x,y) = k(x^2 + y^2)$$
>
> where *k* is a constant. If you were asked to give the following function, what would you write?
>
> $$T(r,\theta) = ?$$

*Figure 2: A problem whose answer tends to distinguish mathematicians from physicists.*

The context of the problem encourages you to think in terms of a particular physical system. Physicists tend to think of *T* as a *physical function* – one that represents the temperature (in whatever units) at a particular point in space (in whatever coordinates). Mathematicians tend to consider *T* as a *mathematical function* – one that represents a particular functional dependence relating a value to a pair of given numbers. The physicist tends to answer that $T(r,\theta) = kr^2$ since they interpret $x^2 + y^2$ physically as the square of the distance from the origin. If *r* and $\theta$ are the polar coordinates corresponding to the rectangular coordinates *x* and *y*, the physicist's answer gets the same value for the temperature at the same physical point in both representations. In other words, the physicist assigns meaning to the variables *x*, *y*, *r*, and *θ* – the geometry of the physical situation relating the variables to one another. Our mathematician, on the other hand, may regard them as *dummy* variables denoting two independent variables--so, (*r*, *θ*) or (*x,y*) don't have any geometric meaning constraining their relationship. The mathematician focuses on the grammar of the expression rather than the physical meaning. The function as defined instructs one to square the two independent variables, add them, and multiply the result by *k*. The result should therefore be $T(r,\theta) = k(r^2 + \theta^2)$.[1]

---

[1] A "shibboleth" is a word or phrase that can be used to distinguish members of a group from non-members.

Typically, a physicist will be upset at the mathematician's result. You might hear, "You can't add $r^2$ and $\theta^2$! They have different units!" The mathematician is likely to be upset at the physicist's result. You might hear, "You can't change the functional dependence without changing the name of the symbol! You have to write

$$T(x,y) = S(r,\theta) = kr^2. \qquad (1)$$

To which the physicist might respond, "You can't write that the temperature equals the entropy! That will be too confusing." (Physicists often use *S* to represent entropy.)

Of course, there are times when physicists pay careful attention to functional form – for example, when considering the transformation from a Lagrangian to a Hamiltonian or between thermodynamic potentials. (Zia et al., 2009) But once they are experts, the context suffices to alert physicists to the need to pay attention to mathematical grammatical element of functional form.

### *Novices often pay attention only to mathematical grammar*

Not only do mathematicians often focus on the mathematical grammar of an equation while ignoring physical meaning; novice students often do the same. A few years ago, one of us (EFR) gave the example shown in *Figure 3* to his second-semester class in algebra-based physics.

> A very small charge *q* is placed at a point $\vec{r}$ somewhere in space. Hidden in the region are a number of electrical charges. The placing of the charge *q* does not result in any change in the position of the hidden charges. The charge *q* feels a force, *F*. We conclude that there is an electric field at the point $\vec{r}$ that has the value *E = F/q*.
>
> If the charge *q* were replaced by a charge –3*q*, then the electric field at the point $\vec{r}$ would be
>
>     a) Equal to –*E*
>     b) Equal to *E*
>     c) Equal to –*E*/3
>     d) Equal to *E*/3
>     e) Equal to some other value not given here.
>     f) Cannot be determined from the information given.

*Figure 3: A quiz problem that students often misinterpret.*

---

[1] It is to be noted that we are exaggerating the roles of the mathematician and physicist thinking to illustrate the point that one could have starkly different interpretation of the same equation. We are not implying that a real world mathematician would be not think of Cartesian to polar coordinate transformations at all if presented with this problem.

The instructor had discussed the definition of electric field extensively and he had made it "quite clear" that the strength of the electric field was independent of the test charge. Yet on this quiz, more than half of nearly 200 students thought that the answer should be (c). Upon discussing it with the class after the quiz, it became clear that, although many of them could quote the result "the electric field is independent of the test charge," most of the students answering incorrectly had not thought to access that knowledge. They looked at the equation $E = F/q$ and treated the problem as a problem in mathematical grammar. If $E = F/q$, then $F/(-3q) = -E/3$. In this situation, however, $F$ is the Coulomb force on the test charge due to interaction with the source charge and thus increases proportionally as the test charge is increased, resulting in the *same* field as before. Novices' propensity to not integrate these algebraic symbols with physical meaning leaves them vulnerable to such errors. Of course, leaving the dependence of $F$ on $q$ implicit has a role in lending coherence to the novice error on this problem-- but that does not undermine our claim that students reason in terms of mathematical grammar, missing clues to a different interpretation given by physical meaning. Writing $F(q)$ might help a few students realize that the algebraic symbol $F$ is a function of $q$, but not that these symbols in physics equations stand for real physical quantities.

### *The Photoelectric Effect Equation: Interpreting equations in the context of a physical situation can build in implicit conditions*

A third example comes from the modern physics section of the calculus-based physics class for engineers. As usual, the instructor (EFR) tried to "twist" student expectations a bit by making small but important modifications to standard problems. The problem shown in Figure 4 was a surprise to some of the students but the results were more of a surprise to the instructor.

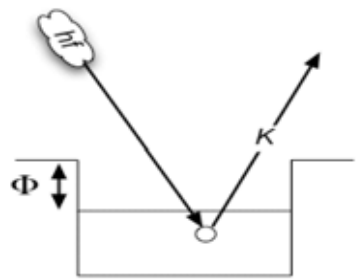

*Figure 4: A problem where the equation hides some physics.*

This one is pretty easy if you keep the physics to the fore. A longer wavelength corresponds to a smaller frequency so the photons have *less* energy after the modification. If the photons did not have enough energy to knock out an electron originally, they would certainly not have enough if we reduce their energy further. Yet nearly a quarter of the class responded that after the modification, there would be electrons knocked out. Their reasoning relied on the Einstein photoelectric effect equation:

$$eV_0 = hf - \Phi \tag{2}$$

where $e$ is the charge on the electron, $V_0$ is the stopping potential, $f$ is the frequency of the photon, and $\Phi$ is the work function of the metal. Their reasoning went as follows: "If $hf - \Phi$ is zero before the change, then it will not be zero after the change. Therefore, since it's not zero there will be electrons knocked out."

This reasoning highlights the fact that Eq. (2) is not the correct equation. There is a hidden Heaviside (theta) function that corresponds to the statement: Do not use this equation unless the right hand side, which corresponds to the maximum kinetic energy of the knocked out electron, is a positive number. A more mathematically correct equation is

$$eV_0 = (hf - \Phi)\, \theta(hf - \Phi) \qquad (3)$$

Typically, expert physicists don't bother to be explicit about this theta function. They simply "check the physics" to see that the energy is positive before applying the equation.

These examples demonstrate that physicists (as well as other scientists and engineers) often use ancillary physical knowledge —often implicit, tacit, or unstated— in their application of mathematics to physical systems. Interestingly enough, a similar idea turns out to be valuable to linguists trying to understand how we put meaning to words – semantics.

## III. Some Relevant Elements of Cognitive Semantics

To understand how we make sense of the language of mathematics in the context of physics, lets consider what is known about how people make sense of language in the context of daily life. Many cognitive linguists have studied semantics – how we make meaning with language – for the past 100 years, and they have made considerable progress in the past 25. Although the research community has not entirely come to an overarching synthesis, they have many ideas that are valuable in helping us make sense of how we make meaning with math in physics. We offer an exceedingly brief summary of a rich and complex subject, selecting those elements that are particularly relevant. We draw heavily on the work of (Lakoff & Johnson 1980), (Lakoff & Nunez 2001), (Langacker 1987), (Fauconnier 1985, 1997), (Turner 1991), (Fauconnier & Turner 2003), (Evans 2006), and (Evans & Green 2006).

**Fundamental Principles**

In their overview of modern cognitive semantics, Evans and Green identify four core ideas (Evans & Green 2006, p. 157):

- *Encyclopedic knowledge*: Ancillary knowledge is critical to determining the meaning of words.
- *Conceptualization*: Meaning is constructed dynamically.
- *Embodied cognition:* The meaning of words is grounded in physical experience.
- *Conceptual grounding*: Semantic structure expresses conceptual structure.

   *Encyclopedic knowledge:*
   ***Ancillary knowledge is critical in the creation of meaning.***

The principle of encyclopedic meaning implies that we understanding the meaning of words not in terms of terse definitions provided in the dictionary but in reference to a contextual web of concepts (represented by other words) that are themselves understood on the basis of still other concepts. Take the example of the word "hypotenuse," defined as the longest side of a right-angled triangle. But, to understand the meaning of "hypotenuse", one needs to understand the meaning of a triangle, "sides" of a triangle, right angle, and the idea of longest. These in turn require conceptualizing a plane, shapes on a plane, lines, angles, length, and so on. In other words, understanding and conceptualizing "hypotenuse" relies on a number of ancillary concepts, which in turn rely on a background of other concepts *in an ever-expanding web of encyclopedic knowledge*.

This idea of encyclopedic knowledge has been developed under various frameworks such as frame-semantics (Fillmore 1976), domains (Langacker 1987), and mental spaces (Turner 1991). The idea common to the various perspectives is that we can model knowledge as consisting of a very large number of elements that are highly interconnected, much like the world-wide-web. At any given moment, only part of the network is active, depending on the present context and the history of that particular network. The meaning of a word in a particular utterance is then determined by what part of that complex web of knowledge is accessed by that particular utterance of that word. Modern cognitive linguists argue compellingly for these complex links in the structure and processing of meaning:

> *…a number of scholars…have presented persuasive arguments for the view that words in human language are never represented independently of context . Instead, these linguists argue that words are always understood with respect to **frames** or **domains** of experience…. a frame or domain represents a schematisation of experience (a knowledge structure), which is represented at the conceptual level and held in long-term memory, and which relates elements and entities associated with a particular culturally-embedded science, situation or event from human experience….words (and grammatical constructions) are relativised to frames and domains so that the 'meaning' associated with a particular word (or grammatical construction) cannot be understood independently of the frame with which it is associated.* (Evans & Green 2006, p. 211)

The meaning associated with a word or concept could also extend to include the knowledge we have about or in relation to the concept via our haptic, visual, auditory, olfactory, and other senses. In *Foundations of Cognitive Grammar,* Langacker (1987, p. 154) argues that cognitively, the meaning of the concept [BANANA] is not restricted to its linguistic specification. Rather the meaning of [BANANA] encompasses its associations within the non-linguistics domains of space (or visual domain), color, and taste/smell sensations as well as cultural knowledge about bananas as edible, about banana trees, where they grow, etc.

At first look, the meaning of mathematical equations could seem terse in the dictionary sense. What does it mean to know the meaning of an equation such as $y = mx + b$? In the strictest sense, it is a statement defining the value of the quantity "$y$" in terms of the sum of "$b$" and the product of "$m$" and "$x$".

We argue that mathematics intentionally adopts such a minimalist view. What you know about a mathematical quantity is specified as precisely and as minimally as possible (with axioms), and only that knowledge is to be used in constructing new knowledge.[1]

Though written mathematics can be terse and precise, we do tend to rely more than just a "dictionary meaning" in how we use and understand mathematics. For most mathematicians (and even high school students) the equation $y = mx + b$ carries more meaning than the literal relation between four symbols. With our knowledge about labeling conventions, $x$ and $y$ are interpreted as variables capable of taking on many different values, while $m$ and $b$ are interpreted as constants. With this addition, the equation takes on the meaning of a relation between the independent variable ($x$) and the dependent variable ($y$). Additionally, the constancy of $m$ implies that the equation refers to a straight line. The constants now take on ad-ditional mathematical meaning: *m* as the slope of the line and *b* as the intercept on the y-axis, bringing in ideas from graphs.  Thus, the meaning of the equation, understood within the domains of mathematical conventions, straight lines, and graphs is much richer then the strict "definition" expressing the symbol "*y*" in terms of other symbols.

---

[1] Although this is the ideal, many mathematicians argue that there are subtle issues that prevent this ideal from ever being realized, even in principle. (Goldstein 2006).

When we use math in physics our rich knowledge about the physical system is additionally brought to bear in interpreting the mathematical meaning as illustrated by the examples in section II.  As another example, think about the following equation from physics $v = v_0 + at$, where *v* is the velocity of an object, $v_0$ is the initial velocity, *a* is the constant acceleration of the object, and *t* is the time.  The equation is identical in grammatical structure to the mathematical equation above.  But in the context of physics, it is connected to an even richer network of ideas involving motion, speeds, and rates of change.  To understand this equation is to understand its relation to all these stores of knowledge.  In other words, the meaning of equations in physics is encyclopedic just like the meaning of a word such as "banana" is encyclopedic.

> *Conceptualization:*
> *Meaning is constructed dynamically.*

One of the implications of the idea of encyclopedic knowledge system is that the meaning of entities not fixed but is dynamically determined based on the specific contextual clues.  Semanticists see linguistic units as prompts for the construction of meaning, or, *conceptualization.*  As described by Evans and Green (200X, p. 162), conceptualization is, " a dynamic process whereby linguistic units serve as prompts for an array of conceptual operations and the recruitment of background knowledge.  It follows from this view, that meaning is a process rather than a discrete "thing" that can be "packaged" by language.  Meaning construction draws upon encyclopaedic knowledge…and involves **inferencing strategies** that relate to different aspects of conceptual structure, organization and packaging..."

If utterances provide access to nodes in the network of knowledge, different parts of which are active in different moments, then the idea of encyclopedic meaning implies that the meaning of an utterance would depend on the local activity of the network at any given moment.  It is in this sense that the meaning is dynamic.  The contexts of a particular utterance – local (what is the conversation about, with whom, etc.) and global (personal histories, social and institutional affordances, etc.) – affect the activity of the knowledge network and in turn, the conceptualization at any given moment.  Thus, the meaning of an utterance is *not* predetermined but is constructed in the moment.  Take the example of the word "safe" in following sentences in the context of a child playing on the beach (Evans & Green 2006, p.161):

> (1) The child is safe
> (2) The beach is safe

The first sentence refers to the child as free from harm.  The second sentence, though identical in structure to the first, does not imply that the beach is free from harm, rather it implies that the beach cannot cause harm.  The senses of "safe" in the two sentences (and the properties they assign to child and beach) are slightly different depending on what the utterance is referring to.  "In order to understand what the speaker means, we draw upon our encyclopedic knowledge relating to children [and beaches] and our knowledge relating to what it means to be safe. We then 'construct' a meaning by 'selecting' a meaning that is appropriate in the context of the utterance." (Evans & Green, 2006, p. 161-162).

In a different context, for example, while referring to a beach threatened by property developers, the conceptualization and the meaning constructed for the second utterance could be quite different.

In the light of this, we can revisit the example of the imaginary physicist and mathematician arguing about whether $T(r,\theta)$ should be "$r^2 + \theta^2$" or "$r^2$". We can now understand the mathematician as attaching meaning to the equation within the domain of simple functions and variables, while the physicist is additionally interpreting the equation within the domains of coordinate systems adding physical meaning to the variables.

This idea ties in nicely with the *resources framework* being developed by the Maryland Physics Education Research Group for analyzing student thinking in physics. (Hammer 2000, Redish 2004, Hammer, Elby, Scherr, & Redish 2004) In this framework, cognitive resources (tightly bound cognitive knowledge structures) are recruited (activated) by associations (through spreading activation) and contexts. The explication of particular inferential strategies by the linguists provide us tools to help us begin to understand the mechanism of how contexts and associational structures recruit/activate particular resources.

> *Conceptual grounding:*
> *Semantic structure expresses conceptual structure.*

To some extent, the cognitive dynamics of meaning construction are expressed in the structure with which we express or communicate that meaning. Modern cognitive linguists argue that the conceptualization in a moment (and correspondingly, the conceptual structure in that moment) is closely bound to the semantic structure. Thus, changes in semantic structure reflect changes in conceptualization. It is important to note here that "semantic structure relates not just to words but to all linguistic units. A linguistic unit might be a word like a *cat*, a bound morpheme such as *-er,* as in *driver* or *teacher*, or indeed a larger conventional pattern, like the structure of an active sentence or a passive sentence." (Evans & Green, 2006, p. 159). Evans and Green argue that even though the objective information or meaning provided by the active sentence ("William Shakespeare wrote Romeo and Juliet") is identical to that provided by the passive sentence ("Romeo and Juliet was written by William Shakespeare"), the difference in the semantic structure reflects a difference in conceptualization and conceptual structure. In the active sentence, we are focusing on the *actor* (William Shakespeare, in this case) while the passive formulation draws greater focus to the *participant undergoing the action* (the play, Romeo and Juliet, in this case). The difference in focus could imply a different organization of the knowledge network, which could in turn lead to differences in how we access subsequent domains of knowledge.

Talmy (2000) suggests that the semantic representation of concepts takes place through the interaction of two systems: a conceptual structuring system and a conceptual content system. The former is highly schematic while the meaning associated with the conceptual content system is rich and highly detailed. While semantic structure must to some extent reflect conceptual structure, the relationship is complex and indirect.

In physics, this could mean that how we express ideas matters in determining what conceptualizations are activated. Mathematically, the following two equations representing Newton's second law convey identical information:

$$\boldsymbol{F} = m\boldsymbol{a}, \qquad (1) \text{ and}$$
$$\boldsymbol{a} = \boldsymbol{F}/m, \qquad (2)$$

where ***F*** is the total force on an object, *m* is the mass of the object, and ***a*** is the acceleration of the object. But the conceptualization of the two equations could be quite different. Equation (1) focuses on the force as the product of the mass and acceleration, often leading students to think of it as the *definition* of force. The equal sign is typically understood as connecting two identical entities. Equation (2), on the other hand, encourages the activation of a causal relationship: acceleration as resulting from the forces acting on an object. The equals sign can now carry a very different meaning of connecting effect to cause.

> ***Embodied cognition:***
> ***The meaning of words is grounded in physical experience.***

While encyclopedic knowledge relates to the topology of the knowledge network and cognitive processes on that network, embodied cognition refers to the interaction of complex cognitive functions with basic sensorimotor routines. If concepts are understood only within a network of other concepts, how then are the first concepts derived? The thesis of embodied cognition states that ultimately our conceptual system is grounded in our interaction with the physical world: How we construe abstract meaning might be constrained by and often derived from our very concrete experiences in the physical world.

Embodied cognition is not a reference to the cognitive activity that is obviously involved in performing sensorimotor activities. The idea is that (a) our close sensorimotor interactions with the external world strongly influence the structure and development of higher cognitive facilities, and (b) the cognitive routines involved in performing basic physical actions are involved in higher-order abstract reasoning.

There are abundant examples in everyday language and conceptualization where meaning cannot be understood without deriving orientation from our bodily existence in the three dimensional world that we experience. Lakoff & Johnson (2004) discuss extensive examples of many spatial orientation concepts and metaphors such as "up," "down," "front," and "back" that are tied closely to our spatial experiences: "... Thus UP is not understood purely in its own terms but emerges from the collection of constantly performed motor functions having to do with our erect position relative to the gravitational field that we live in." (Lakoff & Johnson, 2004, p. 56-57).

This then forms the basis of structuring and understanding other concepts. Thus metaphorical statements such as: "I'm feeling *up*" or "He is really *low* these days" are conceptualized on the basis of physically drooping when one in depressed or in a negative emotional state.

Lakoff and Johnson (2004) point out how our bodily experiences with physical objects form the basis of conceptualizing emotions such as anger, love and events such as inflation in terms of discrete entities or physical substances. Such conceptualization forms the basis on which we use and understand statements such as, "Inflation is backing us into a corner," and "You've got too much hostility in you." Many of these conceptualizations are so ingrained and automatic that we do not even realize the metaphorical nature of them in everyday use.

The embodiment of cognition and meaning construction is what enables us to break the circularity that we referred to in our example of the dictionary, where words and concepts are seemingly defined in terms of yet other words and concepts that one might not know the meaning of. The point is that a dictionary does not provide definitions of words in terms of other words in the dictionary. Its circular chain *implicitly expects* that at some point you will come across words that you learned in a more direct and perceptual context than from a dictionary and whose meaning you can discern *without* further referring to the dictionary.

An infant begins to learn language by direct perceptual experience. Its parents identify multiple objects that are clearly different from each other with the same word, "chair." Part of the process is epigenetic – hard-wired to permit us to learn given the appropriate experiences.[1] A toddler does not have to be taught by its parents to bind the multiple perceptions it receives when it holds its bottle (touch, sight, and taste) into the perception of a single object. But it does have to be taught the word to use to make that bottle appear when the child is hungry. That association of the word with the perceived object is the part on which a dictionary definition, in the end, relies.

The grounding of conceptualization in physical experience and actions also extends to higher cognitive processes such as mathematical reasoning. Lakoff and Nunez (2001) argue that our understanding of many mathematical concepts relies on everyday ideas such as spatial orientations, groupings, bodily motion, physical containment, and object manipulations (such as rotating and stretching): "Mathematical ideas are often grounded in everyday experience. Many mathematical ideas are ways of mathematizing ordinary ideas, as when the idea of a derivative mathematizes the ordinary idea of instantaneous change." (p. 29). Thus, the mathematics of set theory can be understood as grounded in our physical experience with containers and collections of objects, and conceptualizing the arithmetic of complex numbers "makes use of the everyday concept of rotation." (p. 29). In other words, many of the sophisticated ideas and formulations in mathematics are intricately entwined with the physicality of our being.

In the end, all of our complex concepts must eventually come down to direct perceptual experience. This seems an almost unbelievable result, given the complexity and abstractness of concepts we deal with everyday. Some cognitive linguists have begun to develop possible pathways by which those direct perceptions can be elaborated, refined, and extended to create complex abstractions. They include metaphor (Lakoff & Johnson 2004), polysemy (Evans 2006), and blending (Fauconnier & Turner 2003).

**Meaning creates communication stability: Interpreting partial or corrupted text.**

One of the clear bits of evidence supporting the complex view of meaning making outlined in the synthesis by Evans and Green (2006), is the ease with which we make sense of partial or corrupted speech and text. Everyone is likely to recall examples when someone is speaking and makes a slip of the tongue (a verbal typographical error) such as omitting a critical "not" or slipping in the wrong name, but no one has any trouble in getting the intended meaning. Indeed, the listeners may be unaware of the error until it is reviewed on a recording. We have no trouble interpreting the verses of Lewis Carroll's "Jabberwocky", the second verse of which is quoted below, despite the presence of many nonsense words. (Carroll 1871)

> Beware the Jabberwock, my son!
> The jaws that bite, the claws that catch!
> Beware the Jubjub bird, and shun
> The frumious Bandersnatch!

Despite the fact that we have no idea what a "vorpal" sword is or what a "Tumtum tree" might look like (both from later in the poem) we make sense of it. Tenniel's illustration of the Jabberwock brings in a visual non-dictionary-based element that allows the entire poem to become realistic and interpretable.

---

[1] Joaquin Füster refers to such inherited structures as *phyletic memory* – memory we have as a genetic record of our evolution as a species. (Füster 1999).

A major goal of building expertise in physics (or, we suspect, in any other complex knowledge structure) is creating a complex web of meaning – an encyclopedic knowledge base that adds stability and the ability to recover from error.

In the following section we apply the principles of cognitive linguistics to students' reasoning in physics to argue that sophisticated use of mathematics in physics involves connecting the mathematical equations to the rich network of knowledge and intuition about the real world, not just at the level of variables but at the level of the interpretive basics of the mathematical and physical processes. Our example contrasts the explanations of two students who can both solve the physics problems presented to them. The difference in the quality of their use and interpretation of equations can be understood in terms of the encyclopedic network of knowledge that they activate in association with these equations.

## Meaning Making with Equations: contrasting two students

Understanding an equation in physics is not limited to connecting the symbols to physical variables and being able to perform the operations in that equation. An important component is being able to connect the mathematical operations in the equation to their physical meaning and integrating the equation with its implications in the physical world. In this section, we illustrate the differences in ways that meaning could be attached to an equation by analyzing excerpts of clinical interviews (AG was the interviewer) with two students, Pat and Alex[1], in an introductory physics class. The excerpt focuses on the students' understanding of the equation for the velocity of an object moving with a constant acceleration: $v=v_0 + at$, where $v$ is the velocity of the object at time $t$, $v_0$ is the velocity of the object at $t = 0$, and $a$ is the constant acceleration. Though both students can use the equation satisfactorily for solving problems, but the 'encyclopedic meaning' ascribed to the equation by Pat is different than that for Alex.

When Alex is asked to explain the velocity equation, she focuses on the meaning of the symbols:

> AG: Here's an equation; OK, you've probably seen it…
>
> Alex: Yeah.
>
> AG: Right. So suppose you had to explain this equation to a friend from class; how would you go about doing that?
>
> Alex: Umm…OK… Well… umm… I guess, first of all, well, it's the equation for velocity. Umm… *pause* well, I would… I would tell them that it's uh… I mean, it's the integral of acceleration, the derivative of *furrows brow* position, right? So, that's how they could figure it out… uh… I don't know. I don't really *laugh* I'm not too sure what else I would say about it. You can find the velocity. Like, I guess it's interesting because you can find the velocity at any time if you have the initial velocity, the acceleration, time and…

---

[1] Pat and Alex are gender-neutral pseudonyms and throughout the paper we will refer to them by the feminine pronoun "her." We intentionally choose not to identify the gender of our participants so as to maintain the focus on the substance on their reasoning and to preemptively rule out any correlations between their reasoning and their gender.

Her response suggests that Alex thinks about the equation as something that allows one to calculate the velocity of an object at any moment. She does refer to the velocity as the derivative of position and integral of acceleration but her comment does not reflect if those mathematical operations connecting velocity to position and velocity to acceleration have any deeper physical meaning for her.

Pat also refers to the physical meaning of the symbols in the equation, but her explanation is not limited to that. Her explanation seems to dip into knowledge about motion, units, and processes of change:

> AG: Ok. So here's probably an equation that you have seen before, right? And um, if you were to explain this equation to a friend from class, how would you go about explaining this?
>
> Pat: Well, I think that the first thing that you'd need to go over would be the definitions of each variable and what each one means, and I guess to get the intuition part; I'm not quite sure if I would start with dimensional analysis or try to explain each term before that. Because I mean if you look at it from the unit side, it's clear that acceleration times time is a velocity, but it might be easier if you think about, you start from an initial velocity and then the acceleration for a certain period of time increases that or decreases that velocity.

Pat also looks at the physical meaning of each symbol in the equation and how they are connected. She also brings in knowledge about units and how the dimensions in the equation must be consistent between terms. But she lends a deeper meaning to the equation by bringing in additional knowledge about the physical situation of acceleration as a process that changes the initial velocity. Further excerpts of her interview show that, for Pat, the mathematical formulation of the equation is stably connected to the conceptual schema of change, where you start with an initial quantity and then something changes it. The *'at'* term for him reflects the total amount by which the initial value is changed.

Later in the interview, we see that this difference in the interpretation of the equation is also reflected in how Pat and Alex use the equation to solve a problem about differences in the speeds of two balls thrown from a building at the same time with different initial velocities. Alex uses the equation as a tool to generate final velocities given the initial velocity, time, and acceleration. Pat on the other hand uses the equation much more like an expert, reaching the answer without needing to plug in numbers and carry out arithmetic calculations, and she exhibits an expert-like understanding of why the result should be what it is based on the structure of the equation. Pat also spontaneously delves into multiple ways that the problem could be solved and says that if she gets stuck, she would "switch to another one." Alex in response to similar interview prompts seems to think of the answer as just a result of numerical calculations rather than resulting from the structure of the equation and says that one could not have reached the answer without doing the calculations.

## Conclusion

Our point here is that *understanding* equations in physics is not just about learning the mathematical operations. It is about making many connections to stores of knowledge about the physical system; not just at the level of variable-definitions but at level of the bases of the mathematical operations and how those operations connect to physical meaning. Expectations that the mathematics describing a physical system should connect to our intuitive sense about how the physical system behaves can also provide a valuable safety net against possible er-rors in the calculations and help us make sense of new or unfamiliar equations. This perspective, enhanced by the tools developed recently by cognitive semanticists, should help give us new and deeper insights into the use of mathematics in physics and the other sciences.

## Acknowledgements

We are grateful for conversations on the topic of this paper with members of the University of Maryland Physics Education Research Group. This material is based upon work supported by the US National Science Foundation under Awards No. DUE 05-24987 and REC 04-40113. Any opinions, findings, and conclusions or recommendations expressed in this publication are those of the author(s) and do not necessarily reflect the views of the National Science Foundation.